# Detecting ultra-low ozone and hydrogen concentrations with CsPbBr$_3$ microcrystals direct grown on electrodes

Aikaterini Argyrou[1,2], Konstantinos Brintakis[1*], Athanasia Kostopoulou[1*], Emmanouil Gagaoudakis[1,2], Ioanna Demeridou[1,3], Vassilios Binas[1,3,4], George Kiriakidis[1,3,4] and Emmanuel Stratakis[1,3*]

[1]*Foundation for Research and Technology Hellas (FORTH), Institute of Electronic Structure & Laser (IESL), P.O. Box 1385, Heraklion 70013, Crete, Greece*
[2]*University of Crete, Department of Chemistry, 71003 Heraklion, Crete, Greece*
[3]*University of Crete, Department of Physics, 71003 Heraklion, Crete, Greece*
[4]*Crete Center for Quantum Complexity and Nanotechnology, Department of Physics, University of Crete, 71003 Heraklion, Greece*

**ABSTRACT:** Highly stable ozone and hydrogen sensing elements were fabricated based on well- crystalline rounded cube-shaped CsPbBr$_3$ submicron crystals, synthesized by a facile solution process performed under ambient conditions. It is shown that such elements demonstrate enhanced room temperature gas sensing ability compared to the previously reported metal halide and oxide-based ones. Electrical measurements performed on these sensing components revealed high response to ultra-low ozone and hydrogen concentrations, namely 4 ppb and 5 ppm respectively, as well as an impressive repeatability of the sensing behavior even after a few months of storage in ambient conditions. Both ozone and hydrogen detecting sensors were self-powered, i.e. they do not require the use of UV or heating external stimuli, and exhibited fast detection and short restoration times. These attractive properties along with the simple synthesis conditions could provide an easy, efficient and low-cost potential technology for the realization of future gas sensing devices.

The great demand to detect a broad range of combustible, toxic and explosive gases has led to a significant development of gas sensing technologies. Particularly, gas sensors exhibit a wide range of either commercial or industrial applications including air-quality monitoring,[1] public safety[2,3] and agriculture.[4] The diversity in gas sensing technologies arises from the variety of the developed materials which interact with target gases.[5] The most common sensing materials are the metal oxide semiconductors due to their low-cost production and high sensitivity, however, they usually operate at high temperatures and exhibit low selectivity.[6,7] In these materials, the detection of gases is based on the resistance changes due to reactions that occur at the surface of the sensing element.[8]

Among a wide range of gas sensors, those that detect ozone and hydrogen hold a prominent role due to the hazardous effects of both gases in ambient air. Ozone is mainly found in the stratosphere where it is known for its protective role against the UV radiation. In lower concentrations, ozone is also detected in ground-level and is considered as a common air pollutant. As a strong oxidizing gas, ozone is also used in several applications such as water disinfection[9] and food processing[10]. However, it is well known that in high concentrations in the ambient atmosphere, ozone gas is harmful to human health.[11] Short-term exposure to ozone is related to reduced lung function, increased respiratory infections, asthma crisis and more, while long-term exposure has be linked to permanent lung damage or even death.[12,13] In order to expand the utilization of ozone sensors, it is necessary to develop both sensitive and inexpensive sensor devices for the continuous real-time monitoring of its concentration.

On the contrary, hydrogen is the most abundant chemical element in the universe. At room temperature and atmospheric pressure, pure hydrogen exists as a diatomic gas in concentrations less than 1 ppm.[14] Hydrogen gas is extensively used in many industrial processes such as in the synthesis of ammonia[15] and methanol,[16] the production of rocket fuels[17] and metal refining[18] and is the most attractive candidate for future fuel and energy carrier.[19] Hydrogen gas is colorless, odorless, tasteless and highly explosive when its concentration to the air exceeds the 40000 ppm.[6, 20] However, hydrogen gas is very susceptible to leakage due to its small molecular volume, and therefore, the need for its detection and low concentration monitoring is essential.

Perovskites crystals have attracted the interest of the scientific community mainly as materials for energy conversion and storage.[21,22] Recently, due to the fact that their optical and electronic properties can be reversibly changed after their exposure to external stimuli, perovskites have been also emerged as new sensing elements.[23–25] The reversibility of those properties in the presence and absence of an external stimuli allowed perovskites to be used for the detection of many factors such as humidity,[26] a variety of gases,[27–30] solvents[31] and explosives[32]. Additionally, the high tolerance of all-inorganic perovskite nanocrystals to environmental factors such as moisture and oxygen compared to their hybrid organic-inorganic counterparts makes them prominent competitors against other semiconducting materials. [33–36]

Halide perovskites were lately introduced as sensing elements with great potential in ozone and hydrogen sensing technology. In particular, a hybrid metal halide $CH_3NH_3Pb_{3-x}Cl_x$ thin film was investigated as both ozone and hydrogen sensing material with a demonstrated detection limit of 5 ppb and 10 ppm respectively.[37,38] However, the complexity of the fabrication process, including high annealing temperatures and spin coating techniques, as well as the sensitive to oxygen and moisture organic nature of the film limits its potential applications. Besides the organic-inorganic halides, all-inorganic perovskite nanocrystals have also been reported for ozone detection. Recently, our group fabricated a ligand-free $CsPbBr_3$ gas sensor by a solution-based process under controlled, inert atmosphere, exhibiting high performance to concentrations below 200 ppb and remarkable stability over a three-month period.[39]

Herein, we first report on ultrasensitive and highly stable $CsPbBr_3$-based ozone and hydrogen sensors, fabricated under ambient conditions and operating in room temperature. It is striking that the ambient synthesis environment favored the growth of rounded cube-shaped crystals (RC), exhibiting superior sensing performance compared to that we have reported for $CsPbBr_3$ cubes synthesized in inert Ar atmosphere. Electrical current intensity measurements carried out at room temperature revealed excellent sensitivity, reversibility and repeatability of the sensing elements.

**EXPERIMENTAL SECTION**

**Fabrication of the sensor:** A precursor solution consisted of 0.4 mmoles $PbBr_2$ (trace metals basis, 99.999%, Sigma-Aldrich) and 0.4 mmoles CsBr (anhydrous, 99.999%, Sigma-Aldrich) dissolved in 10 ml N,N-dimethylformamide (DMF) (anhydrous, 99.8%, Alfa Aesar) was prepared by following an already published chemical protocol under the protective atmosphere of an argon-filled Glovebox (MBRAUN, UNILab).[39] Then, the sealed vial was removed from Glovebox and ligand-free $CsPbBr_3$ crystals were fabricated under ambient conditions by a facile re-precipitation method. 15 μl of precursor solution and 15 μl of toluene (99.8%, Sigma-Aldrich) were deposited on commercially available interdigitated platinum electrodes (IDEs, 10μm bands/gaps, Metrohm) on glass substrate. At this stage, the color of the sample changed instantly to yellow, indicating the fast nucleation and the growth of crystals. The substrate was left to dry under ambient conditions for an hour.

**Characterization of the sensor:** The surface morphology of $CsPbBr_3$ crystals was characterized by using Scanning Electron Microscopy (SEM, JEOL 7000) operating at 20 kV. The elemental analysis of the as prepared samples was estimated by Energy Dispersive X-Ray Spectroscopy (EDS, INCA PentaFET-x3). The crystal structure was investigated by X-Ray Diffraction (XRD, Rigaku D/MAX2000H) in a 2θ = 5 to 70° angle range with 0.01° step and a 2 s step time, using a monochromatic CuKa radiation source. The absorbance spectrum of the $CsPbBr_3$ crystals was measured by UV-Visible spectroscopy (Perkin Elmer Lambda 950 UV/Vis/NIR) in the wavelength range of 250 to 800 nm. The Micro-Photoluminescence (μ-PL) spectrum of $CsPbBr_3$ crystals was investigated by a home-made set-up at room temperature, in a backscattering geometry with a Diode-Pumped Solid-State (DPSS) Laser Nd:YAG 473 nm as excitation source.

The gas sensing measurements were performed at room temperature, in a home-made gas sensing chamber, which allowed the electrical characterization of the sensors in a controlled atmosphere. The gas sensing assembly consists of certified gas suppliers connected with mass flow controllers and a stainless chamber which was initially evacuated down to $10^{-3}$ mbar. For the investigation of the ozone sensing capability of $CsPbBr_3$ crystals, a commercial ozone analyzer (Thermo Electron Corporation, Model 49i) was used to supply and record well defined ozone concentrations which are inserted in the chamber with a 500 sccm (standard cm³/min) flow. Electrical current measurements as a function of time were carried out by an electrometer (Keithley 6517A) by applying a constant voltage of 1 V. The sensing process was initiated by exposing the sensor to ozone gas for 10 min, and then, the sensor's recovery was followed by replacing the ozone gas with synthetic air for 20 min. The sample was exposed to different ozone concentrations ranging from 2620 to 4 ppb. The hydrogen sensing capability of $CsPbBr_3$ crystals were examined in the same set-up by uncoupling the synthetic air supplier and the ozone analyzer and connecting certified pure nitrogen and hydrogen gas sources. The sensing process was initiated by introducing hydrogen gas for 3.5 min.

Subsequently, the sensor was recovered by introducing in the chamber pure nitrogen for 5 min. The sensing performance of the sample was studied under different $H_2$ concentrations ranging from 100 to 1 ppm. During both experimental procedures the pressure in the chamber was kept constant and equal to 600 mbar. The sensitivity S of the sensors was calculated according to the equation $S\% = \frac{I_{max} - I_{min}}{I_{min}} \times 100$, where $I_{max}$ and $I_{min}$ denote the maximum and minimum current values in the presence and absence of the test gas respectively.

**RESULTS AND DISCUSSION**

Ligand-free $CsPbBr_3$ crystals were directly grown onto commercially available interdigitated electrodes under ambient conditions via a simple re-precipitation technique. The morphology of $CsPbBr_3$ crystals can be well described as a cube with rounded edges (**Figure 1a**) exhibiting a size distribution of 0.55 μm ± 0.38 μm (**Figure S1**). The rounded edges across the cubes may be attributed to the water content present in toluene during the crystallization process. It is well reported that the exposure of lead halide perovskites in traces of water during the re-precipitation method affects the final morphology of the crystals and depending on the water concentration, it can lead to the formation of various shapes ranging from square-like shaped crystals to nanowires.[40]

The XRD pattern of the rounded perovskite cubes exhibited sharp diffraction peaks, which correspond to that of the orthorhombic phase of the $CsPbBr_3$ (**Figure S2**). At the same time, some additional low-intensity peaks are also identified, indicating the existence of the secondary tetragonal phase $CsPb_2Br_5$. The formation of this secondary phase with reduced crystal dimensionality (2D) is possibly associated with a water-induced transformation from the $CsPbBr_3$ phase due to synthesis conditions.[40,41] In addition, elemental analysis, performed by EDS, confirmed the atomic ratio of Cs:Pb:Br to be 1.2:1:3.1 (**Figure S3**).

**Ozone sensing capability of the CsPbBr3 rounded cubes:** The ozone sensing properties of $CsPbBr_3$ RCs were evaluated via electrical measurements performed at room temperature. In a typical experiment, the sensor was first placed into a home-made gas-sensing set-up and a voltage of 1 V was applied. The initial current value was almost equal to 3 μA that can attributed to the strong photoresponse of $CsPbBr_3$ rounded cubes to the visible light, which can be also indicated by their sharp absorption band edge at 544 nm (**Figure S4**).[42] Subsequently, the sensor was exposed to ozone gas and as shown in **Figure 1b**, the electrical current instantly increases until a maximum value. Then, the ozone gas is replaced by synthetic air and the sensor recovers within twenty minutes. As shown in the same figure, a similar reversible electrical behavior was noticed upon exposure of the sensor to different ozone concentrations ranging from 2620 to 4 ppb. To further investigate the ability of the sensor to resolve low ozone concentrations with accuracy, the normalized exponential growth curves during the oxidation procedure, for the different ozone concentrations, tested, were plotted in the inset of **Figure 1b**. It can be observed that the sensor is capable to detect and distinguish ultra-low ozone concentrations even at the level of 4 ppb, which is the lowest reported detection limit among other metal halides.[37,39]

What can be further observed from the sensing properties depicted in **Figure 1b** is that as the ozone concentration increases, the corresponding maximum ($I_{max}$) current intensity tends to increase as well, while in the absence of the oxidizing gas the minimum ($I_{min}$) current intensities remain fairly constant indicating the good reversibility of the sensing process (**Figure 1c**). As a consequence of the difference in the $I_{max}$ values, the sensitivity of the sensor is affected by the ozone concentration. In particular, as shown in **Figure 1c**, the calculated sensitivity decreases from 250% for 2620 ppb to approximately 13% for the lowest detection limit of 4 ppb.

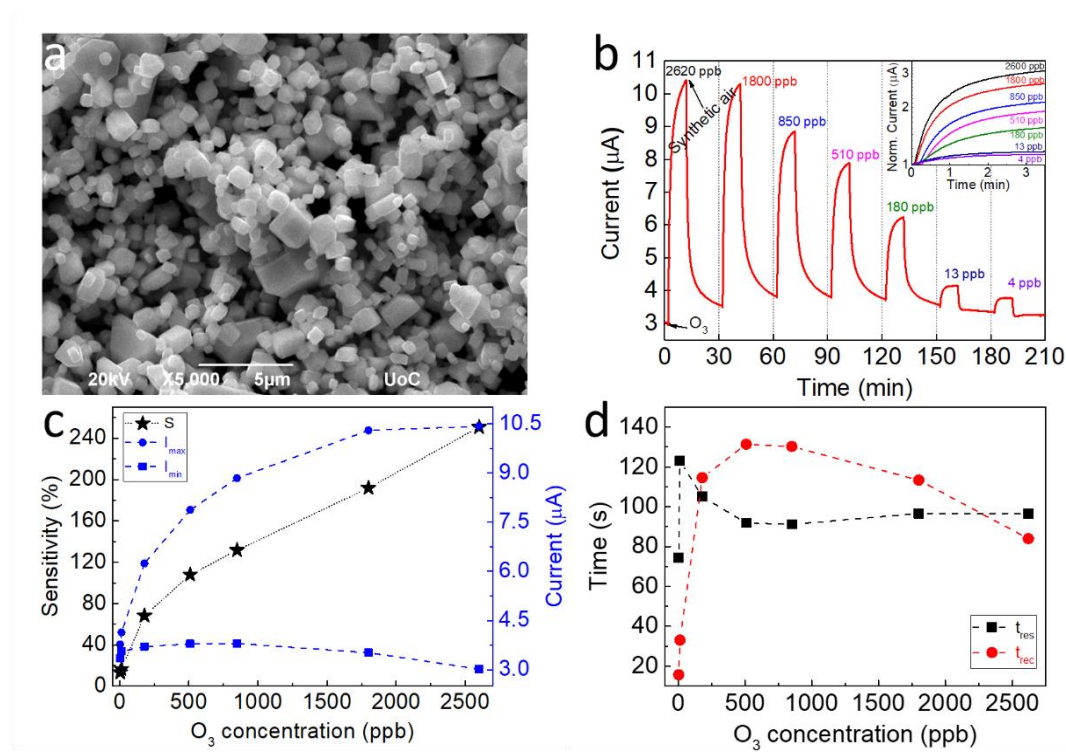

**Figure 1.** (a) SEM image of $CsPbBr_3$ rounded cubes. (b) Electrical response of the sensor following exposure to ozone concentrations ranging from 2620 to 4 ppb; the normalized exponential growth curves during the oxidation process as a function of time (inset); (c) Sensitivity, and (d) response/recovery times as a function of ozone concentration.

Besides the sensitivity, rapid detection and recovery are two important requirements for a promising sensing system. Since the electrical response of the sensing element can be described by an exponential growth and decay for each gas concentration, the corresponding response and recovery times, associated with the growth and decay coefficients can be calculated using **Equation E1** (supporting information). **Figure 1d** presents the different response ($t_{res}$) and recovery ($t_{rec}$) times calculated for the different ozone concentrations.

Despite the common demand in sensing technology to detect at the lowest possible response time, the sensor stability and repeatability are also crucial performance features that determine the quality of a sensor. To examine the repeatability of the fabricated sensors, we have performed four ozone/synthetic air exposure cycles, . In all such cycles the ozone concentration used was kept constant at 510 ppb. As presented in **Figure 2a**, each cycle was characterized by a rise in the sensing current as the ozone was inserted into the chamber accompanied by a current decay to its initial value when the ozone gas was replaced by synthetic air, indicating the high reversibility and stability of the sensing element upon consecutive cycles. The repeatability of the sensing behavior was re-examined after three months of the sensor storage in ambient conditions. In this case, it was noticed that, possibly due to adsorbed oxygen molecules, the sensing currents recorded were higher compared to the initial ones, however the sensing behavior was similar (**Figure 2b**). SEM imaging and EDS analysis confirmed that the RC maintained their morphological and stoichiometric features (**Figure 2c, S5 a and b**). Finally, the XRD spectra analysis revealed that the sensing RC retained their initial orthorhombic crystal structure (**Figure 2d**).

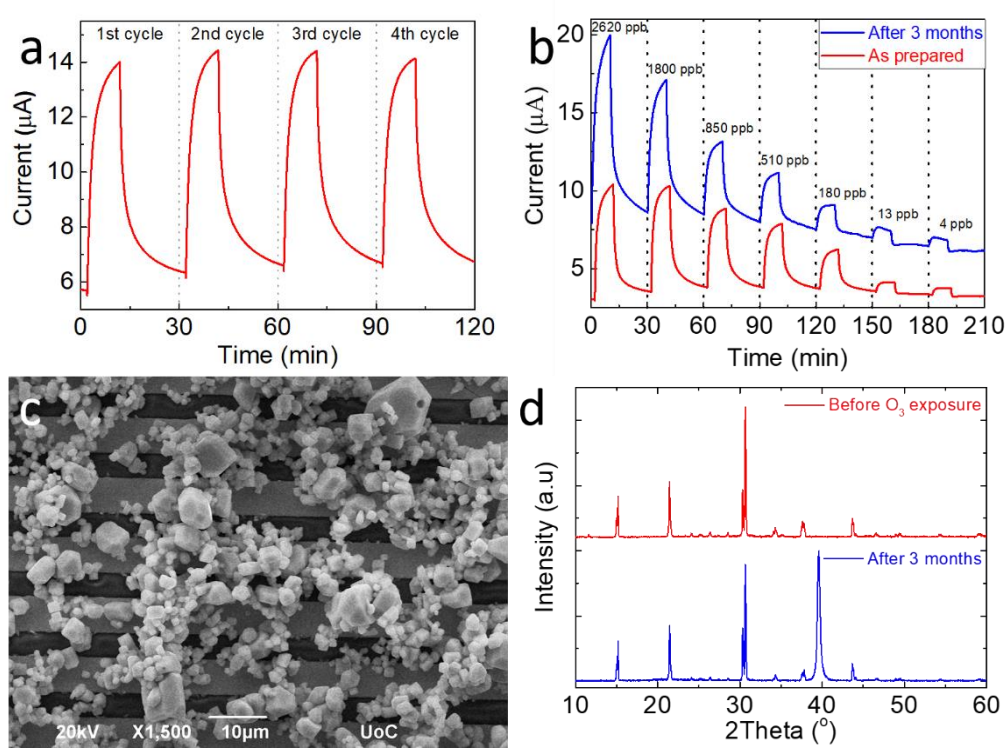

**Figure 2.** (a) Short-term stability of the $CsPbBr_3$ sensor following 4 ozone/synthetic air switching cycles. (b) Comparison of the electrical response of the pristine sensor (red curve) and that measured after three months (blue curve) of its in ambient conditions. (c) SEM image at the end of the sensing process. (d) Comparison of the XRD patterns before (red curve) and after three months (blue curve) of its storage in ambient conditions. The extra peak at ~40° is assigned to the interdigitated Pt electrodes of the substrate used (ICSD 52250).

**Hydrogen sensing properties of $CsPbBr_3$ rounded cubes:** The ability $CsPbBr_3$ RC, to detect hydrogen was also investigated via electrical measurements at room temperature. In this case nitrogen gas has been used as a carrier for the sensor's recovery. The electrical response of the sensor as a function of time for all the hydrogen levels tested is depicted in **Figure 3a**. The sensor was initially exposed to 100 ppm hydrogen gas for 3.5 min, resulting to an instant increase of the current intensity. Subsequently, upon exposure to nitrogen gas for 5 min, the sample was recovered to its initial state. The same procedure was repeated for all hydrogen concentrations ranged from 100 to 1 ppm and similar performance was observed. The corresponding normalized exponential growth curves as a function of time during the hydrogen treatment were well-distinguished, even between the lowest concentrations of 5 and 1 ppm (**Figure 3a inset**). This indicates the sensor's ability to resolve extremely low hydrogen concentrations. As shown in **Figure 3b** the calculated sensitivity values increased from 1.5% at 1 ppm to almost 94% at 100 ppm (**Figure 3b**).

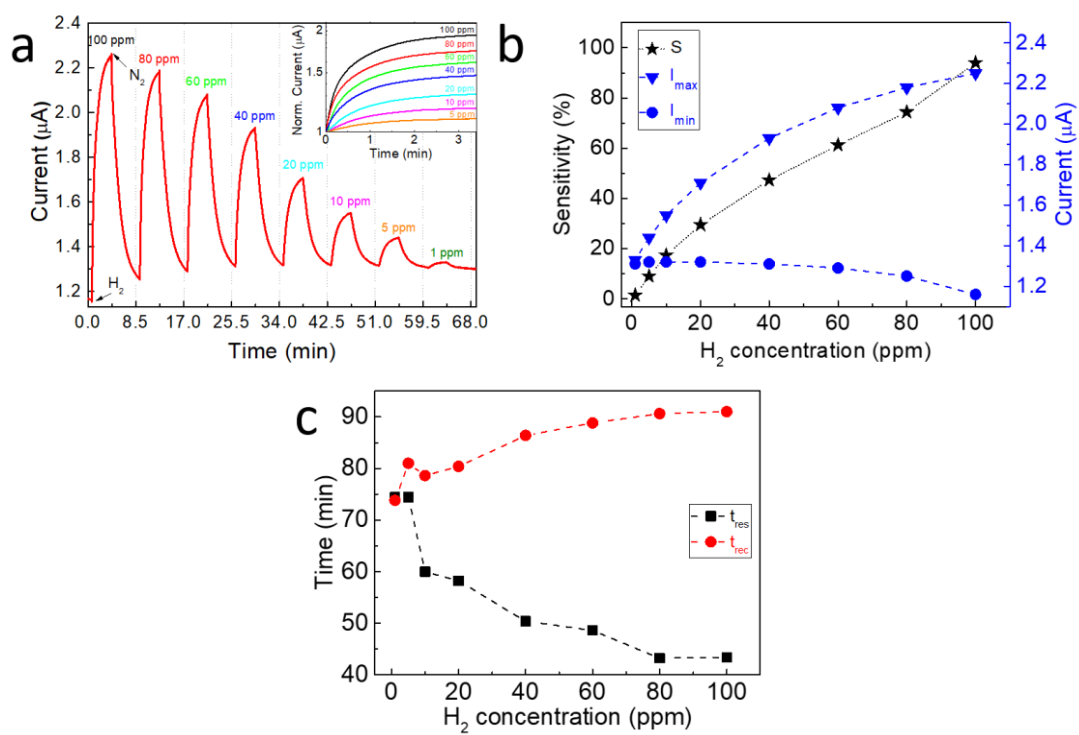

**Figure 3.** (a) Electrical response of CsPbBr$_3$ rounded cubes as a function of time, following exposure to hydrogen concentrations ranged from 100 to 1 ppm. The inset shows the normalized exponential growth curves during the hydrogen treatment as a function of time. (b) Sensitivity and (c) response and recovery times (c) as a function of H$_2$ concentration.

**Figure 3c** presents the progression of response and recovery times as a function of the hydrogen levels tested. It can be observed that the sensor exhibits quick response and short recovery with values ranging between 43 and 75 s and 74 to 91 s, respectively. In particular, at 1 ppm, which is the lowest detection limit of the sensor, the response time is 74.4 s while the recovery time is 73.8 s.

The short-term repeatability of the sensor was also investigated following a series of H$_2$/N$_2$ sensing cycles (**Figure 4a**). It can be observed that the sensor did not fully recover, as the current did not return to its initial value, possibly due to trapped hydrogen molecules. Besides this, the reproducibility of the sensing process was investigated, following the sensor' storage for two months in ambient conditions. **Figure 4b** shows the corresponding performance for different gas concentrations ranging from 100 to 1 ppm. It can be observed that even after a long-term exposure to ambient conditions, the sensor was able to detect ultra-low concentrations of hydrogen. However, the electrical response of the sensor was deteriorated from the first hydrogen treatment. SEM imaging showed that in this degraded state there are no morphological alterations of the crystals following the storage period (**Figure 4c**). Furthermore, the XRD pattern of the sample after the storage process is similar to that of the as prepared one, indicating that no alterations of the crystal structure took place (**Figure 4c**).

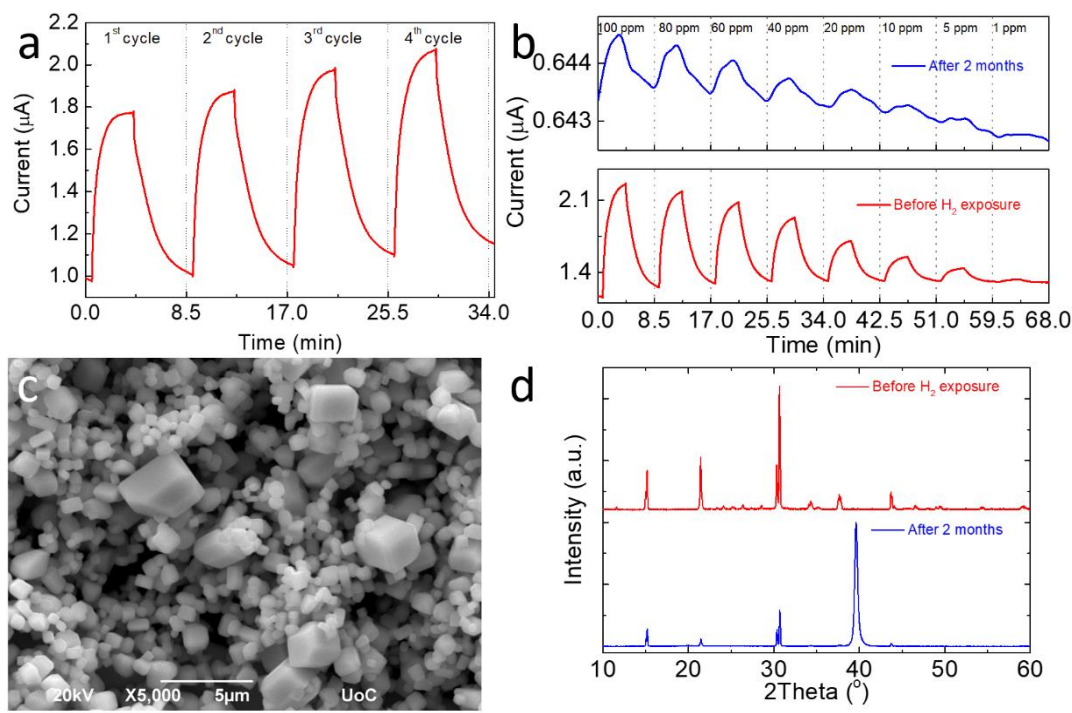

**Figure 4.** (a) Short-term stability after 4 consecutive sensing cycles as a function of time. (b) Comparison of electrical response of the sensor before (lower, red line) and after two months of its storage in ambient conditions (upper, blue line). (c) SEM image at the end of the hydrogen exposure. (d) Comparison of XRD patterns before (upper, red line) and after two months of storage under ambient conditions (lower, blue line). The extra peak at ~40° is assigned to the Pt electrodes of the substrate (ICSD 52250). The decrease of the intensity is attributed to larger uncovered areas of Pt electrodes.

**Proposed sensing mechanisms: Ozone.** It is well known that morphology, size and crystal structure of the sensing elements affect significantly the performance of a gas sensor.[43] Among the different sensing mechanisms that describe the behavior of the gas sensors, the Wolkenstein adsorption model seems to be appropriate in the present case.[44,45]. This model explains the ozone sensing behavior of the well-defined $CsPbBr_3$ nanocubes as well and has been described in our previous study (**Supporting Info S1**).[39] However, it was noticed that rounded cubes exhibit higher electrical response compared to the previously reported well-defined $CsPbBr_3$ cubes (**Figure S6**). The difference between the current intensities could be attributed to the morphology of the sensing elements. Since adsorption is a surface phenomenon, higher surface-to-volume ratio leads to higher adsorption. As shown in **Figure 1a**, the edges of the $CsPbBr_3$ RC are slightly deformed resulting to an increase of their surface area, which means they probably exhibit higher surface-to-volume ratio compared to well-formed $CsPbBr_3$ cubes, which could explain the higher sensitivity of the sensor. Additionally, the improved sensing behavior could also be assigned to the presence of the secondary $CsPb_2Br_5$ phase. The existence of $CsPb_2Br_5$ particles in dual-phase $CsPbBr_3$-$CsPb_2Br_5$ composites affects the electrical properties of $CsPbBr_3$ crystals by decreasing their trap density.[46] As a consequence, the number of non-radiative electron-hole recombinations is decreased, allowing more free charge carriers to participate in charge transport.[47] This hypothesis could be confirmed by the enhanced PL intensity of $CsPbBr_3$ RC compared with pure $CsPbBr_3$ crystals (**Figure S7**).

**Hydrogen.** Cesium lead halide crystals are known for their high tolerance towards defects.[36] Previous reports have shown that halogen vacancies are the most abundant defects in such materials.[48] Halogen vacancies act as major electron traps or recombination centers and mostly appear at the surface of nanocrystals.[49] In our case, the presence of Br vacancies in $CsPbBr_3$ RC can be confirmed by the emission spectrum of the crystals, which exhibit two sub-peaks located at approximately 529 and 540 nm (**Figures S7**) and according to the literature, the peak centered at ~529 nm can be assigned to recombinations involving Br vacancies that lie above the conduction band maximum.[50,51] The sensing mechanism of hydrogen is different to the previous described for ozone, since hydrogen is known as a reducing agent while oxygen is an oxidizing agent and in both sensors the current intensity increases under the gaseous ambient. Consequently, the proposed mechanism is based on trap healing. Hydrogen molecules are adsorbed onto the perovskite surface. Since hydrogen molecule is the smallest size molecule that exists, it is the one which diffuses the most easily in materials. Hence, hydrogen molecules are diffused inside the perovskite crystal structure and fill bromide vacancies which are intrinsically present in the crystal. Since bromide vacancies act as electron traps, their decreased number in presence of hydrogen translates into a higher number of holes that are available for electrical transport, resulting to a current increase.[52]

**Comparing with current state-of-the-art semiconducting elements:** A variety of different semiconducting materials have been suggested for the detection of ozone and hydrogen gases. Comparing the performance for ozone detection of our sensor with the latest studies, the $CsPbBr_3$ rounded cubes (RC) exhibit the best performance among other semiconducting materials since this sensor is able to detect extremely low ozone concentration (4 ppb) at the highest reported sensitivity at room temperature working conditions (**Figure 5a**). The response ($t_{res}$) and recovery ($t_{rec}$) times of $CsPbBr_3$ RC are significantly shorter in duration compared to the already reported halide perovskites and some metal oxides (**Table S1**). On the other hand, materials like Au@$TiO_2$ and NiAl-LDH that exhibit faster response and recovery times, they lack due to their high detection limit along with their poor sensitivity and this confirms the major contribution of this study. [53,54]

Our sensor's hydrogen detection limit at 1 ppm is among the lowest reported to date (**Figure 5b**). Moreover, the sensitivity of $CsPbBr_3$ RC at 5 ppm exceeds the 9%, which is thirty times higher compared to the sensitivity of the $CH_3NH_3PbI_{3-x}Cl_x$ thin film at the same hydrogen concentration.[37] Although, the $t_{res}$ and $t_{rec}$ times are relatively slower compared to some of those that have been reported the recent years for hydrogen sensing at room temperature operating conditions, CsPbBr$_3$ RC can detect extremely low hydrogen concentrations and can be synthesized by an easier and faster process as described above (**Table S2**).

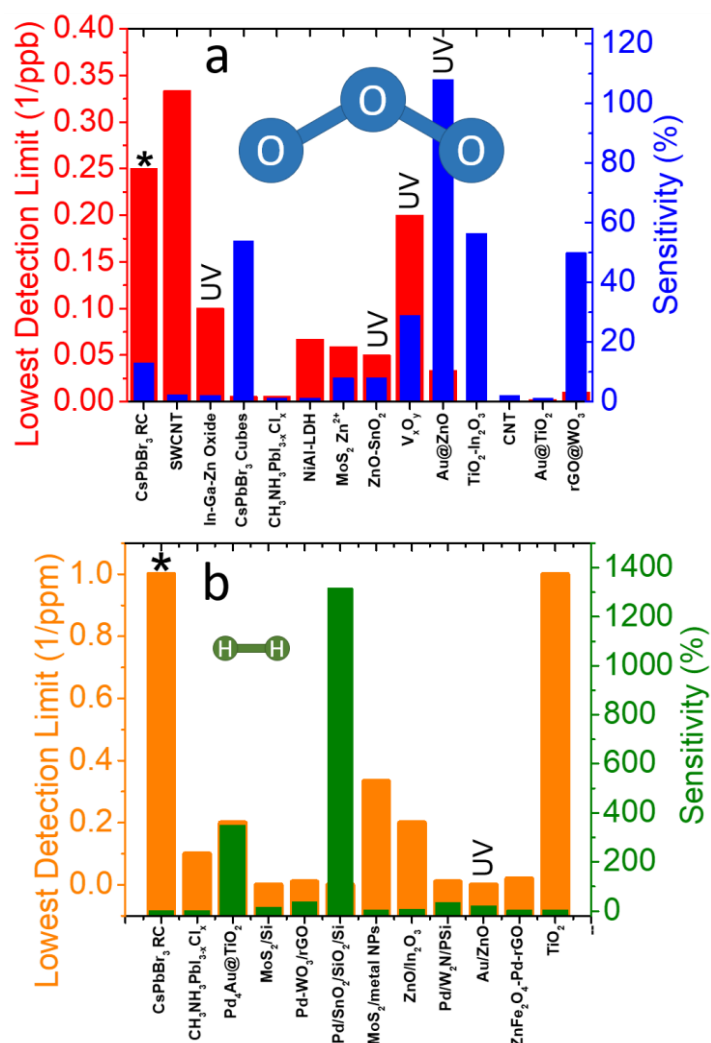

**Figure 5.** Ozone and hydrogen's lowest detection limits (red and orange axes respectively, 1/ppb for better representation) and sensitivities (blue and green axes respectively), comparing other semiconducting sensing elements working at room temperature.[38,39,53-74]. Our results are marked with (*). Insets, schemes of representative ozone and hydrogen molecules taking into account their sizes.

**CONCLUSIONS**

Highly-stable, ligand-free CsPbBr$_3$ rounded cubes have been prepared via a facile solution-based method under ambient conditions. The direct growth of the crystals onto electrodes offers a simpler and cheaper process for perovskite-based elements with enhanced sensing properties. Initially, this method was used for the preparation of an ultrasensitive self-powered ozone sensor (operating without triggering by an external stimuli (UV or heating)) with remarkable stability over time. In addition, its significantly high sensitivity (13% at 4 ppb) compared to those reported in the literature, its excellent repeatability under high $O_3$ concentrations and its quick response and restoration could lead to new opportunities in gas sensing technology. The enhanced sensitivity as well as the ease of the preparation method made $CsPbBr_3$ rounded cubes a good candidate for the detection of other gases. Hence, they were also tested as a hydrogen sensing elements. In this case, the sensor was able to detect fast, extremely low hydrogen concentrations, with sensitivity ranging from 1.5% at 1 ppm $H_2$ concentration to 94% at 100 ppm. Moreover, its detection limit is among the lowest reported compared to other metal halides and oxides at room-operating temperature, providing new opportunities in $H_2$ sensing applications.

## ASSOCIATED CONTENT

**Supporting Information**.

Size distribution, XRD patterns, EDS, absorption and PL spectra, electrical characterization of $CsPbBr_3$ well defined and rounded cubes under different ozone concentrations and comparison of different sensing elements.

## AUTHOR INFORMATION
**Corresponding Authors**

***Konstantinos Brintakis** - *Foundation for Research and Technology Hellas (FORTH), Institute of Electronic Structure & Laser (IESL), P.O. Box 1385, Heraklion 70013, Crete, Greece.*; Email: kbrin@iesl.forth.gr

***Athanasia Kostopoulou** - *Foundation for Research and Technology Hellas (FORTH), Institute of Electronic Structure & Laser (IESL), P.O. Box 1385, Heraklion 70013, Crete, Greece.*; Email: akosto@iesl.forth.gr

***Emmanuel Stratakis** - *Foundation for Research and Technology Hellas (FORTH), Institute of Electronic Structure & Laser (IESL), P.O. Box 1385, Heraklion 70013, Crete, Greece.*; *University of Crete, Department of Physics, 71003 Heraklion, Crete, Greece*; Email: stratak@iesl.forth.gr

## ACKNOWLEDGMENTS

FLAG-ERA Joint Transnational Call 2019 for transnational research projects in synergy with the two FET Flagships Graphene Flagship & Human Brain Project - ERA-NETS 2019b (PeroGaS: MIS 5070514). K.B. acknowledges E.U. H2020 Research and Innovation Program under Grant Agreement N 820677 and Greek State Scholarships Foundation (IKY) through the operational Program «Human Resources Development, Education and Lifelong Learning» in the context of the project "Reinforcement of Postdoctoral Researchers - 2nd Cycle" (MIS-5033021). A.K. acknowledges the Hellenic Foundation for Research and Innovation (HFRI) and the General Secretariat for Research and Technology (GSRT), under Grant Agreement No 1179 funded this project. We would like also to thank Mrs Alexandra

# Supporting Information

# Detecting ultra-low ozone and hydrogen concentrations with CsPbBr$_3$ microcrystals direct grown on electrodes

Aikaterini Argyrou[1,2], Konstantinos Brintakis[1*], Athanasia Kostopoulou[1*], Emmanouil Gagaoudakis[1,2], Ioanna Demeridou[1,3], Vassilios Binas[1,3,4], George Kiriakidis[1,3,4] and Emmanuel Stratakis[1,3*]

[1]*Foundation for Research and Technology Hellas (FORTH), Institute of Electronic Structure & Laser (IESL), P.O. Box 1385, Heraklion 70013, Crete, Greece*
[2]*University of Crete, Department of Chemistry, 71003 Heraklion, Crete, Greece*
[3]*University of Crete, Department of Physics, 71003 Heraklion, Crete, Greece*
[4]*Crete Center for Quantum Complexity and Nanotechnology, Department of Physics, University of Crete, 71003 Heraklion, Greece*

Corresponding authors:

- Konstantinos Brintakis: kbrin@iesl.forth.gr
- Athanasia Kostopoulou: akosto@iesl.forth.gr
- Emmanuel Stratakis: stratak@iesl.forth.gr

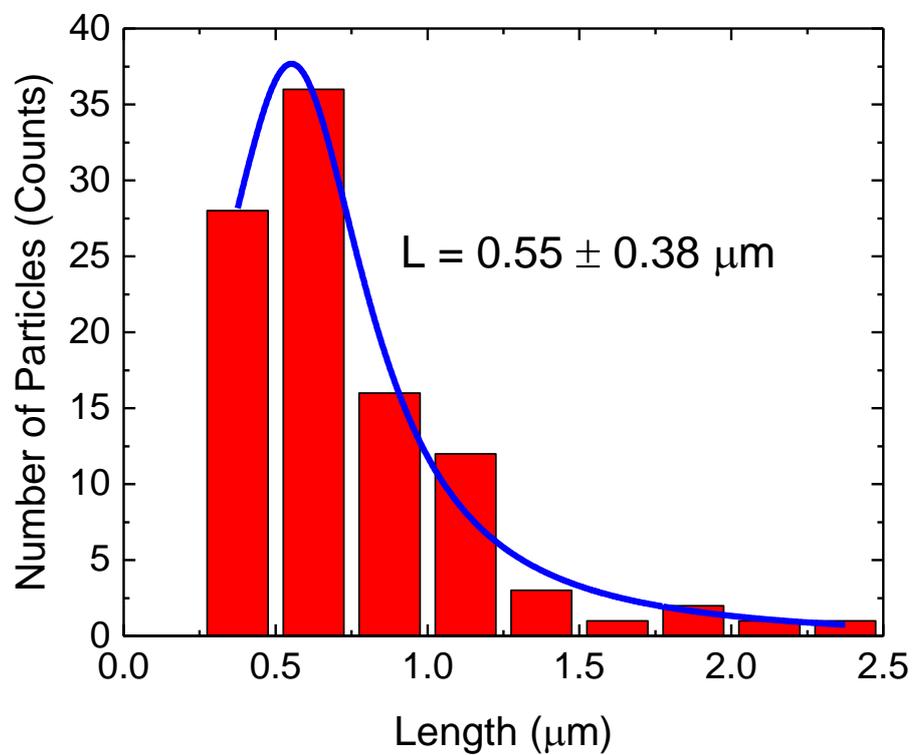

**Figure S1.** Size distribution analysis for ligand-free CsPbBr$_3$ rounded cubes (RC) with an average size of 0.55 ± 0.38 μm.

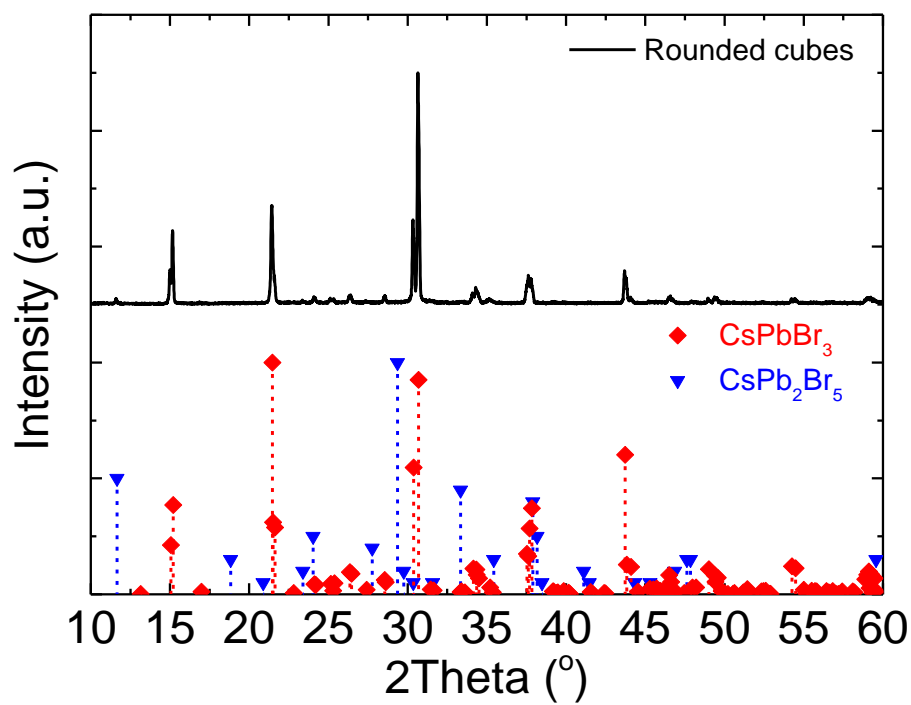

**Figure S2.** XRD pattern of $CsPbBr_3$ rounded cubes before sensing (upper, black line) and reference patterns of the orthorhombic phase $CsPbBr_3$ (lower, red symbol, COD-1533062) and tetragonal phase $CsPb_2Br_5$ (lower, blue symbol, PDF pattern: 025-0211).

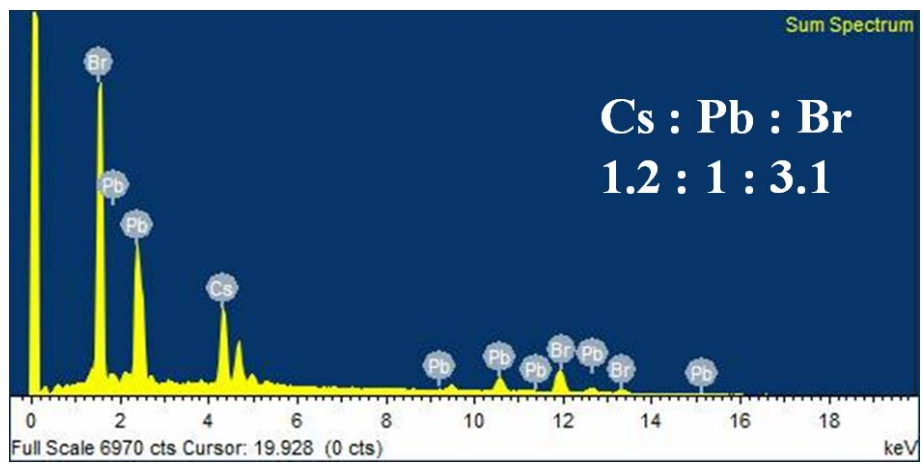

**Figure S3.** EDS spectra of $CsPbBr_3$ rounded cubes before $O_3$ exposure.

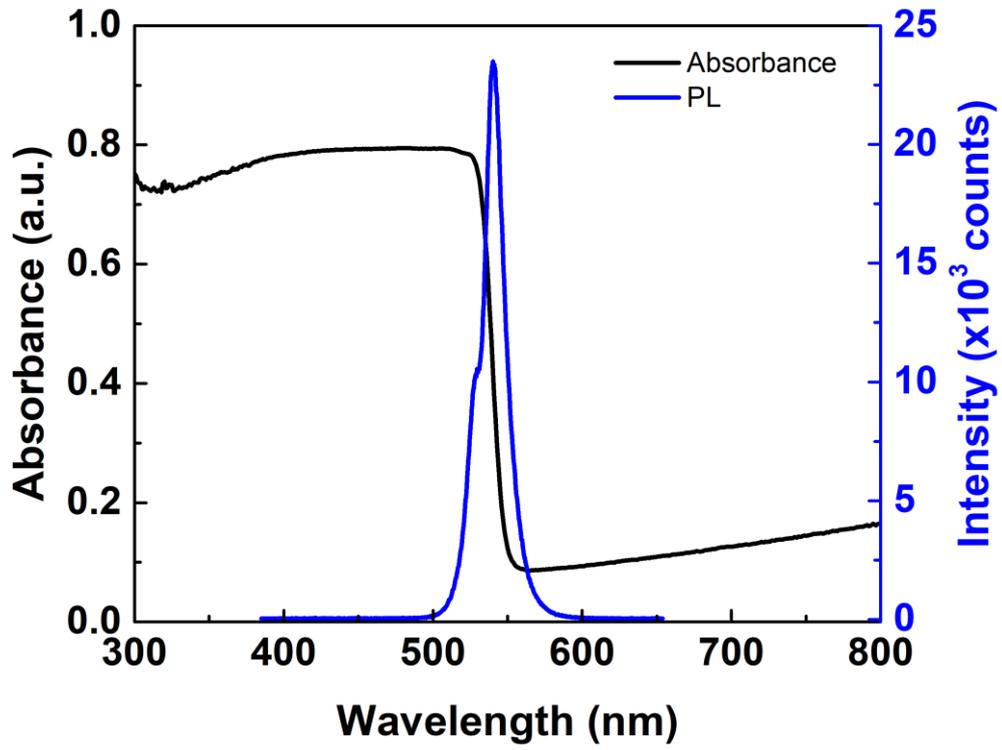

**Figure S4.** Absorption spectrum (black line) and PL (blue line) of CsPbB$_3$ rounded cubes.

**Equation E1:** Exponential fitting where I$_o$ is the initial current value, A$_d$ and A$_g$ are the decay and growth amplitudes respectively and t$_c$ is the time where current reaches its maximum value.

$$I = \begin{cases} I_0 + A_d + A_g \left( e^{-\frac{t_c}{t_{res}}} - e^{-\frac{t}{t_{res}}} \right), & I \leq I_0 \\ I_0 + A_d \, e^{-\frac{t-tc}{t_{rec}}}, & I > I_0 \end{cases}$$

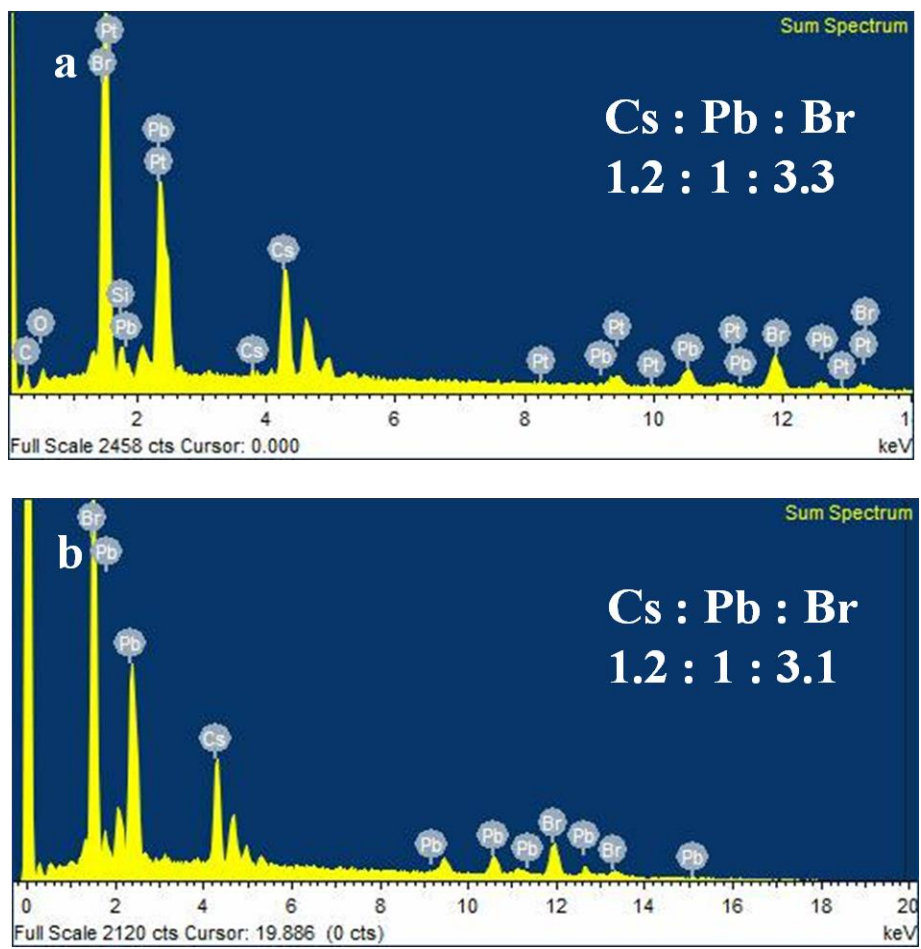

**Figure S5.** (a) EDS spectra of CsPbBr$_3$ rounded cubes after O$_3$ exposure and (b) at the end of the sensing process three months later.

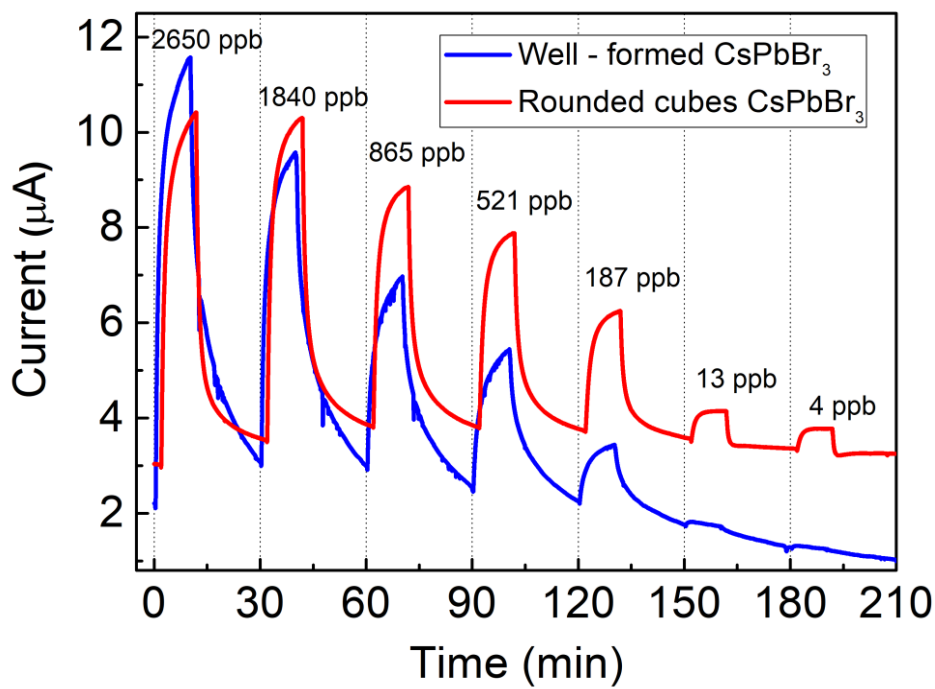

**Figure S6.** Comparison of the current response of the well-formed (blue curve)[1] and rounded (red curve) CsPbBr$_3$ nanocubes as O$_3$ sensing elements.

**S1. Wolkenstein's model:** This model relies on the interaction between the gas molecules and the semiconductor' surface. In particular, the exposure of rounded cubes to ambient air results to the adsorption of oxygen molecules onto their surfaces. Consequently, the oxygen molecules trap electrons from the perovskite's valence band leading to an increase of the hole concentration, and hence, to the formation of a conductive surface layer and a resistive core.[14,15,16] Subsequently, when the sensor is exposed to ozone gas, ozone molecules are adsorbed on the CsPbBr$_3$ surface and act as electron acceptors. As a consequence, the number of holes in CsPbBr$_3$ shell lattice increases, resulting to the decrease of the resistance of CsPbBr$_3$ RC and, hence, an increase in their conductivity.

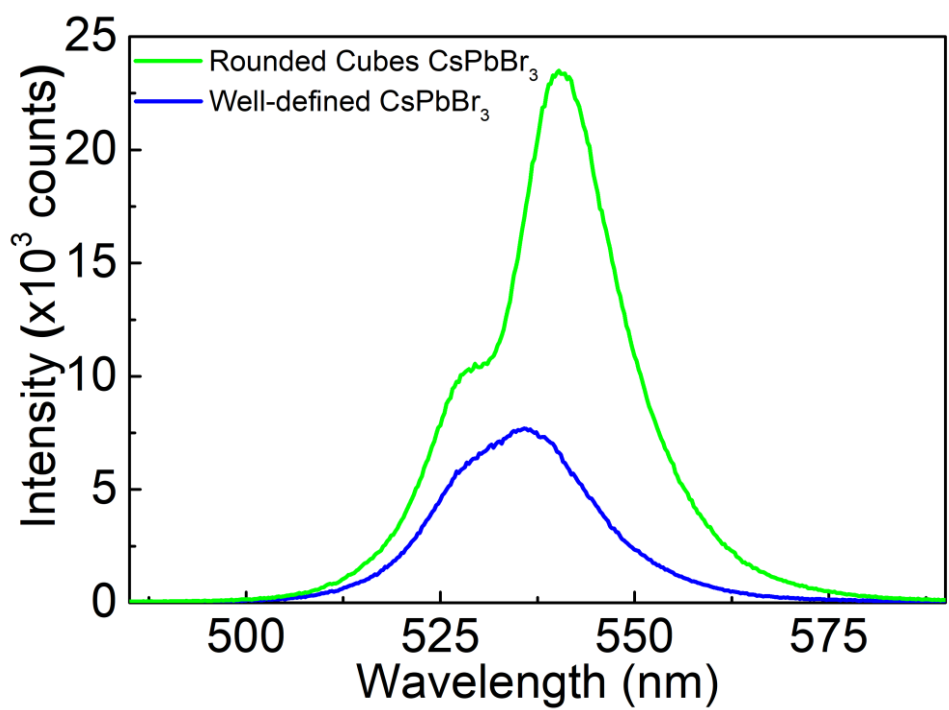

**Figure S7.** PL comparison of well-defined CsPbBr$_3$ cubes (blue line) and CsPbBr$_3$ rounded cubes (green line).

**Table S1.** Comparison of response and recovery times of ozone sensing elements at room temperature working conditions.

| Material | Ozone Concentration (ppb) | Response Time (s) | Recovery Time (s) | Ref. |
|---|---|---|---|---|
| CsPbBr$_3$ RC | 4 | 74.4 | 15.6 | This work |
| CsPbBr$_3$ cubes | 187 | 143 | 320 | 1 |
| CH$_3$NH$_3$PbI$_{3-x}$Cl$_x$ | 180 | 225 | 200 | 2 |
| Au@TiO$_2$ | 500 | 3 | 5 | 3 |
| V$_x$O$_y$ | 5 | 120 | 143 | 4 |
| In-Ga-ZnO | 10 | - | - | 5 |
| TiO$_2$-In$_2$O$_3$ | 2000 | 115 | 143 | 6 |
| ZnO-SnO$_2$ | 20 | 19 | 30 | 7 |
| NiAl-LDH | 15 | 4 | 4 | 8 |
| MoS$_2$Zn$^{2+}$ | 600 | 5.5 | 10.1 | 9 |
| rGO@WO$_3$ | 100 | - | - | 10 |

| Material | | | | |
|---|---|---|---|---|
| SWCNT | 3 | <40 | - | 11 |
| Au@ZnO | 30 | 13 | 29 | 12 |
| CNT | 200 | 2148 | 1902 | 13 |

**Table S2.** Comparison of response and recovery times of hydrogen sensing elements at room temperature working conditions.

| Material | Hydrogen Concentration (ppm) | Response Time (s) | Recovery Time (s) | Ref. |
|---|---|---|---|---|
| $CsPbBr_3$ RC | 1 | 74.4 | 73.8 | This work |
| $CH_3NH_3PbI_{3-x}Cl_x$ | 10 | 38.52 | 61.08 | 17 |
| $Pd_4Au@TiO_2$ | 5 | 21 | - | 18 |
| $MoS_2/Si$ | 5000 | 105 | 443.5 | 19 |
| $Pd-WO_3/rGO$ | 100 | 52 | 155 | 20 |
| $Pd/SnO_2/SiO_2/Si$ | 500 | 145 | 17 | 21 |
| $MoS_2$/metal NPs | 3 | >900 | - | 22 |
| $ZnO/In_2O_3$ | 5 | ~480 | ~480 | 23 |
| $Pd/W_2N/PSi$ | 100 | 24 | 38 | 24 |
| Au/ZnO | 5 | 4 | 24 | 25 |
| $ZnFe_2O_4$-Pd-rGO | 50 | 35 | 4 | 26 |
| $TiO_2$ | 1 | 9 | <200 | 27 |
| OMT | 100 | 85 | 198 | 28 |